\documentclass[12pt]{iopart}
\usepackage{iopams,epsf,psfig}

\def\nothing#1{}

\newcommand{\half}{\mbox{$\textstyle \frac{1}{2}$}}

\newcommand{\roothalf}{\mbox{$\textstyle \frac{1}{\sqrt{2}}$}}

\begin{document}
\title[Quantum-state transformations]
{On optimum Hamiltonians for state transformations}

\author[Brody and Hook]{Dorje~C~Brody and Daniel~W~Hook}

\address{Blackett Laboratory, Imperial College, London SW7
2BZ, UK}

\begin{abstract}
For a prescribed pair of quantum states $|\psi_I\rangle$ and
$|\psi_F\rangle$ we establish an elementary derivation of the
optimum Hamiltonian, under constraints on its eigenvalues, that
generates the unitary transformation $|\psi_I\rangle
\to|\psi_F\rangle$ in the shortest duration. The derivation is
geometric in character and does not rely on variational calculus.
\end{abstract}

\submitto{\JPA}

\vskip 0.5cm

Recently Carlini, {\it et al}.~\cite{carlini} considered the
following problem: What is the optimum choice of the Hamiltonian,
under a given set of constraints, such that the transformation
between a designated pair of quantum states is achieved in the
shortest possible time? Evidently this question is of relevance to
the implementation of various algorithms in quantum computation
(see, e.g., \cite{carlini,boscain} and references cited therein).
Two specific examples for the constraints on the Hamiltonian are
considered in \cite{carlini}, and the optimum solutions are obtained
using the method of variational calculus.

The purpose of this paper is to show that analogous results can be
obtained more directly by use of the symmetry properties of the
quantum state space, hence avoiding the use of variational calculus.
Our approach is closely related to the idea considered in
\cite{brody}, where an elementary derivation is provided for the
minimum time required to transform one quantum state into another
for a given Hamiltonian. The idea here is to reverse the argument to
find the optimum choice of the Hamiltonian that achieves the
transformation in the minimum time.

Consider a Hilbert space ${\mathcal H}^{n+1}$ of dimension $n+1$,
and assume that an arbitrary pair of initial and final states
$|\psi_I\rangle$ and $|\psi_F\rangle$ are specified. The task is to
find the Hamiltonian $H$ on ${\mathcal H}^{n+1}$ that generates the
unitary transformation $|\psi_I\rangle \to|\psi_F\rangle={\rm
e}^{{\rm i} H\tau/\hbar}|\psi_I\rangle$ in shortest possible time
$\tau$. Clearly, if the differences between the eigenvalues of the
Hamiltonian are allowed to take large values, then the value of
$\tau$ can be made very small. This is because the `speed' of a
unitary evolution is proportional to the energy uncertainty (the
so-called Anandan-Aharonov relation~\cite{anandan}). As a
consequence, if the differences between eigenvalues can be made
large, the energy uncertainty can also be made large. Hence we
impose the constraint that the difference of the largest and the
smallest eigenvalues of $H$ be bounded by a constant.

\begin{figure}[t]
 {\centerline{\hspace{-3.3cm}\psfig{file=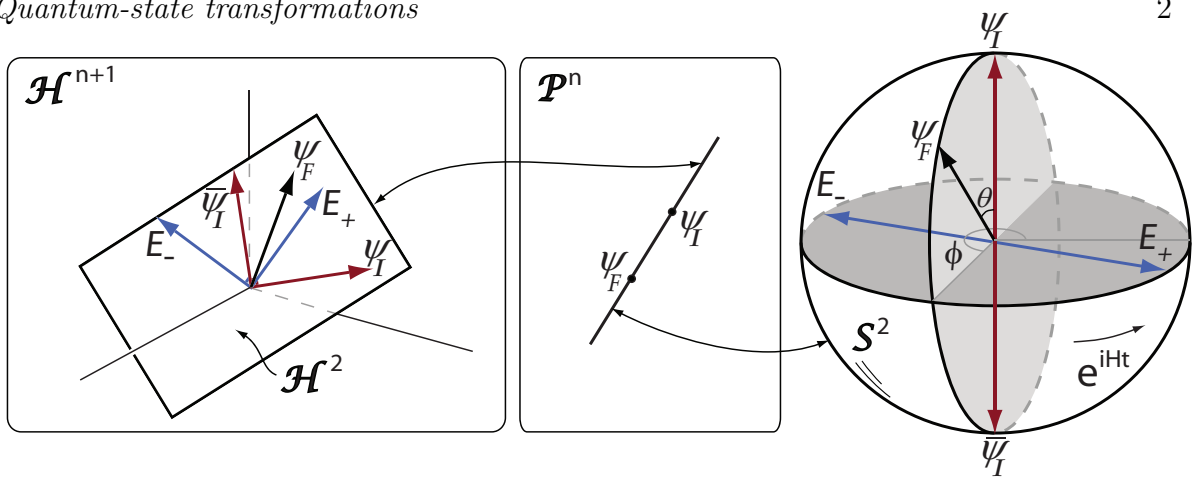,width=12cm,angle=0}}}
 \vskip 0.2cm
   \caption{\textit{Optimum state transformation}. In Hilbert space
   ${\mathcal H}^{n+1}$ one wishes to unitarily transform the initial
   state $|\psi_I\rangle$ into the final state $|\psi_F\rangle$ in
   the
   shortest possible time. In ${\mathcal H}^{n+1}$ there is a unique
   two-plane ${\mathcal H}^{2}$ that contains the two endpoints of
   the vectors $|\psi_I\rangle$ and $|\psi_F\rangle$, and the
   origin. In projective terms this plane corresponds to a complex
   projective line in the space ${\mathcal P}^n$ of pure states, and
   the two vectors $|\psi_I\rangle$ and $|\psi_F\rangle$ determine a
   pair of points on this line. The geodesic curve that joins these
   two points lies on this complex line, which in real terms is just a
   two-sphere $S^2$. Given a pair of points $\psi_I$ and $\psi_F$ on
   $S^2$ there is a unique great circle arc passing through these
   points. The most expedient transformation $|\psi_I\rangle\to
   |\psi_F\rangle$ is thus obtained by the rotation of $S^2$ around
   the axis that is orthogonal to the hemispherical plane containing
   $\psi_I$ and $\psi_F$. The axis of rotation, in particular,
   corresponds to a pair of orthogonal states $|E_\pm\rangle$. The
   Hamiltonian that generates this rotation therefore takes the form
   $H=\lambda_+ |E_+\rangle\langle E_+| + \lambda_- |E_-\rangle
   \langle E_-|$ for a pair of real parameters $\lambda_\pm$.
   \label{fig:1}
   }
\end{figure}

For the analysis of a problem of this kind it is useful to work
directly with the space of rays through the origin of ${\mathcal
H}^{n+1}$. This is just the complex projective space ${\mathcal
P}^n$ of dimension $n$; each ray $|\varphi\rangle\in{\mathcal
H}^{n+1}$ then corresponds to a point $\varphi\in{\mathcal P}^n$.
Thus ${\mathcal P}^n$ can be thought of as the space of directions
in ${\mathcal H}^{n+1}$. Now given a pair of points
$\psi_I,\psi_F\in{\mathcal P}^n$ corresponding to the states
$|\psi_I\rangle$ and $|\psi_F\rangle$ in ${\mathcal H}^{n+1}$ we can
join these two points by a line. The points on this line correspond
to all possible linear superpositions of the states $|\psi_I\rangle$
and $|\psi_F\rangle$. That is, the (complex) line in ${\mathcal
P}^n$ corresponds to the two-dimensional subspace of ${\mathcal
H}^{n+1}$ spanned by the two vectors $|\psi_I\rangle$ and
$|\psi_F\rangle$ (see Figure~\ref{fig:1}). In real terms the complex
line in ${\mathcal P}^n$ corresponds to a two-sphere (the so-called
Bloch sphere) $S^2$, and the two states thus correspond to a pair of
points on the surface of this two-sphere.

%%%%%%%%%%%%%%%%%%%%%%%%%%%
%%%%%%%%%%%%%%%%%%%%%%%%%%%%

It is evident that there is a unique geodesic curve on $S^2$ that
joins $\psi_I$ and $\psi_F$; this is just the great circle arc
passing through these two points (cf. \cite{hughston,brody2}).
Therefore, the unitary transformation that takes the state $|\psi_I
\rangle$ into $|\psi_F\rangle$ in the smallest possible time is
achieved by a rotation of $S^2$ around the axis such that the
geodesic curve joining $\psi_I$ and $\psi_F$ constitutes the equator
associated with that axis. There are infinitely many other unitary
transformations that achieve the transformation $|\psi_I\rangle
\to|\psi_F\rangle$, however, all these transformations will require
longer times to be realised because the corresponding trajectories
are not geodesic curves.

We thus proceed to determine this axis of rotation. To this end let
us write $|{\bar\psi}_I\rangle$ for the state orthogonal to the
initial state $|\psi_I\rangle$ that is contained in the
two-dimensional span of the initial and final states in ${\mathcal
H}^{n+1}$ (i.e.~the antipodal point on $S^2$). Then the final state
$|\psi_F\rangle$ can be written in the form
\begin{eqnarray}
|\psi_F\rangle=\cos\half\theta|\psi_I\rangle + {\rm e}^{{\rm i}
(\phi+\pi/2)}\sin\half\theta|{\bar\psi}_{I}\rangle. \label{eq:1}
\end{eqnarray}
Since both $|\psi_I\rangle$ and $|\psi_F\rangle$ are prespecified,
the values of the two parameters $\theta,\phi$ are known. Our
objective now is to find the axis defined by a pair of antipodal
points on $S^2$ for which $|\psi_I\rangle$ and $|\psi_F\rangle$ lie
on the equator (see Figure~\ref{fig:1}). Since $|\psi_I\rangle$ and
$|{\bar\psi}_I\rangle$ lie on the equator associated with the
$(E_{+},E_{-})$-axis, conversely the two states $|E_{+}\rangle,
|E_{-}\rangle$ lie on the equator associated with the
$(\psi_I,{\bar\psi}_I)$-axis. Hence these states can be expressed as
equal superpositions of $|\psi_I\rangle$ and $|{\bar\psi}_I\rangle$:
\begin{eqnarray}
|E_+\rangle = \roothalf \left( |\psi_I\rangle + {\rm e}^{{\rm
i}\phi} |{\bar\psi}_I\rangle \right) \quad {\rm and} \quad
|E_-\rangle = \roothalf \left( |\psi_I\rangle - {\rm e}^{{\rm
i}\phi} |{\bar\psi}_I\rangle \right). \label{eq:2}
\end{eqnarray}
Solving (\ref{eq:1}) for $|{\bar\psi}_I\rangle$ and substituting the
result into (\ref{eq:2}) we obtain
\begin{eqnarray}
|E_+\rangle = \frac{1}{\sqrt{2}}\left[\left(1+{\rm i}\frac{\cos
\half\theta}{\sin\half\theta}\right)|\psi_I\rangle-\frac{{\rm
i}}{\sin \half\theta}|\psi_F\rangle\right] \label{eq:3}
\end{eqnarray}
and
\begin{eqnarray}
|E_-\rangle = \frac{1}{\sqrt{2}}\left[\left(1-{\rm i}\frac{\cos
\half\theta}{\sin\half\theta}\right)|\psi_I\rangle+\frac{{\rm
i}}{\sin \half\theta}|\psi_F\rangle\right]. \label{eq:4}
\end{eqnarray}
These states thus determine the axis of rotation that we are
seeking.

Now the unitary rotation that gives rise to the rotation of the two
sphere about the axis $|E_+\rangle$ and $|E_-\rangle$ is generated
by the Hamiltonian
\begin{eqnarray}
H = \lambda_+ |E_+\rangle\langle E_+| + \lambda_- |E_-\rangle\langle
E_-|  \label{eq:5}
\end{eqnarray}
for some choice of real parameters $\lambda_+\neq\lambda_-$.
Substituting (\ref{eq:3}) and (\ref{eq:4}) into (\ref{eq:5}) we can
express this Hamiltonian in terms of the two input states:
\begin{eqnarray}
\fl \hspace{1.0cm} H &\hspace{-1.2cm}=& \hspace{-0.6cm}
\frac{\lambda_+ +\lambda_-}{2\sin^2\half \theta} \Big(
|\psi_I\rangle\langle \psi_I|+|\psi_F\rangle\langle\psi_F|\Big)
\nonumber \\ && \hspace{-0.6cm} +\left[\frac{\lambda_+}{2}\left({\rm
i}\frac{1} {\sin\half\theta}
-\frac{\cos\half\theta}{\sin^2\half\theta}\right)
-\frac{\lambda_-}{2}\left({\rm i}\frac{1}{\sin\half
\theta}+\frac{\cos\half\theta} {\sin^2\half\theta}
\right)\right]|\psi_I\rangle\langle\psi_F| \nonumber \\ &&
\hspace{-0.6cm} + \left[\frac{\lambda_+}{2}\left(-{\rm i}\frac{1}
{\sin\half\theta}-\frac{\cos\half\theta}{\sin^2\half\theta}\right)
-\frac{\lambda_-}{2}\left(-{\rm i}\frac{1}{\sin\half
\theta}+\frac{\cos\half\theta} {\sin^2\half\theta}
\right)\right]|\psi_F\rangle\langle\psi_I|. \label{eq:6}
\end{eqnarray}
Because the Hamiltonian in standard quantum mechanics is defined up
to an overall additive constant, without loss of generality we may
set $\lambda_+-\lambda_-=\xi$, and hence,
$\lambda_+=-\lambda_-=\xi/2$, for some real parameter $\xi$. It then
follows at once from (\ref{eq:6}) that
\begin{eqnarray}
H = {\rm i}\xi\frac{1}{2\sin\half\theta}|\psi_I\rangle\langle\psi_F|
-{\rm i}\xi\frac{1}{2\sin\half\theta}|\psi_F\rangle\langle\psi_I|.
\label{eq:7}
\end{eqnarray}
Finally we shall impose the constraint that the difference of the
largest and the smallest eigenvalues (here there are only two) of
the Hamiltonian be given by $2\omega$. Since the eigenvalues of $H$
in (\ref{eq:7}) are $\pm\xi/2\sin\frac{1}{2}\theta$ we have $\omega
= \xi/2\sin\frac{1}{2}\theta$.

More generally, we may consider time dependent Hamiltonians.
However, because of the constraint on the difference of the
eigenvalues, the parameter $\omega$ cannot vary in time. As a
consequence the only time-dependence that can be introduced here is
that associated with the overall magnitude of the Hamiltonian, which
in itself does not affect the dynamics. Letting $h(t)$ denote this
gauge term and ${\bf 1}$ denote the identity operator, the optimum
choice for the Hamiltonian can thus be written as
\begin{eqnarray}
H = {\rm i} \omega|\psi_I\rangle\langle\psi_F| -{\rm i} \omega
|\psi_F\rangle\langle\psi_I| + h(t){\bf 1}. \label{eq:8}
\end{eqnarray}
This is the main result obtained in \cite{carlini}. We emphasise
that this result is obtained here from the symmetry properties of
the quantum state space, essentially only requiring the use of
elementary trigonometry.

As noted above, the time it takes to achieve the transformation
$|\psi_I\rangle \to |\psi_F\rangle$, under the unitary evolution
generated by the Hamiltonian (\ref{eq:8}), can be determined from
the Anandan-Aharonov relation \cite{anandan}, which states that the
`speed' of the evolution of a given quantum state is given by
$2\hbar^{-1}\Delta H$, where $\Delta H$ is the standard deviation of
the Hamiltonian. Note that the energy variance is a constant of
motion. Therefore, calculating the standard deviation of
(\ref{eq:8}) in the state, say, $|\psi_I\rangle$, we deduce that
\begin{eqnarray}
\Delta H=\omega\sin\frac{1}{2} \theta. \label{eq:9}
\end{eqnarray}
On the other hand, the separation of the two states $|\psi_I\rangle$
and $|\psi_F\rangle$ is just the angle $\theta$. We thus find that
\begin{eqnarray}
\tau=\frac{\hbar\theta}{2\omega\sin\frac{1}{2}\theta}.
\label{eq:10}
\end{eqnarray}
Alternatively, the time required for achieving the transformation
can be determined more explicitly as follows. We take the
Hamiltonian (\ref{eq:8}) and use it to calculate the time-dependence
of the state explicitly as
\begin{eqnarray}
|\psi(t)\rangle &=& \left[ \cos\left(\hbar^{-1}\omega
t\sin\half\theta \right) - \frac{\cos\half\theta} {\sin\half
\theta}\sin\left(\hbar^{-1}\omega t\sin\half\theta\right) \right]
|\psi_I\rangle \nonumber \\ && + \frac{1}{\sin\half\theta} \sin
\left(\hbar^{-1} \omega t\sin\half\theta \right) |\psi_F\rangle,
\label{eq:11}
\end{eqnarray}
where $|\psi(0)\rangle=|\psi_I\rangle$. Evidently the coefficient of
$|\psi_I\rangle$ in the state $|\psi(t)\rangle$ first vanishes at
time $t=\hbar\theta/ 2\omega\sin\frac{1}{2}\theta$, while at that
time the coefficient of $|\psi_F\rangle$ becomes unity.

In the foregoing material we have considered the case for which
there is only one constraint on the Hamiltonian, namely, that the
difference of the largest and the smallest eigenvalues be a
constant. In a more realistic setup, however, there can be further
constraints to limit the allowable operations. Although the use of
variational calculus suggested in \cite{carlini} is quite effective
in general, it should be evident that within a given context, the
determination of the optimum Hamiltonian that achieves the desired
transformation simplifies considerably by taking into account the
symmetries of the relevant state space.

\vspace{0.5cm}
\begin{footnotesize}
\noindent DCB acknowledges support from The Royal Society. The
authors thank I.R.C.~Buckley and L.P.~Hughston for useful
discussion.
\end{footnotesize}

\end{document}